\documentclass[prb,aps,twocolumn,showpacs,superscriptaddress,floatfix]{revtex4}
\usepackage{graphicx}% Include figure files
\usepackage{times}
\usepackage{color}
\definecolor{darkblue}{rgb}{0,0.02,0.45}
\usepackage[colorlinks=true,citecolor=darkblue]{hyperref}

\def \vS {{\bf S}}
\def \vR {{\bf R}}

\def \vq {{\bf q}}

\def \mb {\mu_{\rm B}}

\def \va {{\bf a}}
\def \vb {{\bf b}}
\def \vc {{\bf c}}

\begin{document}

\title{Canted antiferromagnetic order and spin dynamics in the honeycomb-lattice $\rm Tb_2Ir_3Ga_9$}

\author{Feng Ye}
\email{yef1@ornl.gov}
\affiliation{Neutron Scattering Division, Oak Ridge National Laboratory,
Oak Ridge, Tennessee 37831, USA}
\author{Zachary~Morgan}
\author{Wei~Tian}
\author{Songxue~Chi}
%\author{Daniel Pajerowski}
\author{Xiaoping~Wang}
\affiliation{Neutron Scattering Division,
Oak Ridge National Laboratory, Oak Ridge, Tennessee 37831, USA}
\author{Michael~E.~Manley}
\author{David~Parker}
\affiliation{Materials Science and Technology Division,
Oak Ridge National Laboratory, Oak Ridge, Tennessee 37831, USA}
\author{Mojammel~A.~Khan}
\altaffiliation{Now at Department of Physics and Astronomy and Department of Chemistry, Johns Hopkins University, Baltimore, MD 21218, USA}
\affiliation{Materials Science Division, Argonne National Laboratory, 9700 South Cass Avenue, Argonne, Illinois 60439, USA}
\author{J.~F.~Mitchell}
\affiliation{Materials Science Division,
Argonne National Laboratory, 9700 South Cass Avenue, Argonne, Illinois 60439, USA}
\author{Randy Fishman}
\email{fishmanrs@ornl.gov}
\affiliation{Material Science and Technology Division,
Oak Ridge National Laboratory, Oak Ridge, Tennessee 37831, USA}

\date{\today}


\begin{abstract}
Single crystal neutron diffraction, inelastic neutron scattering,
bulk magnetization measurements, and first-principles calculations are used to
investigate the magnetic properties of the honeycomb lattice $\rm Tb_2Ir_3Ga_9$.
While the $R\ln2$ magnetic contribution to the low-temperature entropy indicates a
$\rm J_{eff}=1/2$ moment for the lowest-energy crystal-field doublet,
the Tb$^{3+}$ ions form a canted antiferromagnetic structure below 12.5~K.
Due to the Dzyalloshinskii-Moriya interactions, the Tb moments in the $ab$ plane
are slightly canted towards $\vb $ by $6^\circ$ with a canted moment of 1.22\,$\mb $
per formula unit. A minimal $xxz$ spin Hamiltonian is used to simultaneously
fit the spin-wave frequencies along the high symmetry directions and the field
dependence of the magnetization along the three crystallographic axes.
Long-range magnetic interactions for both in-plane and out-of-plane couplings up to the
second nearest neighbors are needed to account for the observed static
and  dynamic properties. The $z$ component of the exchange interactions
between Tb moments are larger than the $x$ and $y$ components.  This compound
also exhibits bond-dependent exchange with negligible nearest exchange
coupling between moments parallel and perpendicular to the 4$f$ orbitals.
Despite the $J_{{\rm eff}}=1/2$ moments, the spin Hamiltonian is denominated by 
a large in-plane anisotropy $K_z \sim -1$\, meV.  DFT calculations
confirm the antiferromagnetic ground state and the substantial inter-plane
coupling at larger Tb-Tb distances.
\end{abstract}

\maketitle
\section{Introduction}

Materials that support a quantum spin liquid (SL) state are of great
interest in condensed-matter physics. On the honeycomb lattice, it is well known that the
Kitaev model produces various two-dimensional
topological SL states \cite{kitaev06,witczak-krempa14, takagi19}.
 Bond-directional anisotropic exchange on a honeycomb lattice frustrates simple collinear
 magnetic order \cite{kitaev06, chaloupka10} in $4d$ and $5d$
transition-metal candidates such as $\alpha$-RuCl$_3$ \cite{plumb14, banerjee16}
and $\rm A_2IrO_3$ (A = Li, Na) \cite{jackeli09,chaloupka10,singh12}, where
strong spin-orbit coupling (SOC) produces $J_{{\rm eff}} = 1/2$ moments. These
systems underscore the recent interest in the honeycomb structural motif.

Decorating the honeycomb lattice with rare-earth ions offers an
alternative to $4d$- and $5d$-based materials. For example, YbMgGaO$_4$, YbCl$_3$, and TbInO$_3$
are proposed quantum SL candidates \cite{Li15, xing20, clark19, kim19} with $J_{{\rm eff}}=1/2$.
Recent theoretical treatments of SOC entanglement in
rare-earth honeycomb magnets motivates further exploration of similar
systems \cite{luo20,jang19}.

A nearly ideal honeycomb lattice of rare-earth ions occurs in the family
$\rm R_2T_3X_9$, where R is a rare-earth element, T is a transition-metal
element, and X is a $p$-block element.
Occupying a large composition space, this family hosts a rich variety of
electronic properties including complex magnetic order for
Dy-based compounds \cite{gorbunov18}, mixed valence in Yb/Ce-based compounds
\cite{gordon96,dhar99,trovarelli99}, and Kondo-lattice behavior for the
Yb-based compounds \cite{okane02,troc07}.

With an orthorhombic crystal structure of the $\rm Y_2Co_3Ga_9$ type
\cite{gladyshevskii92,schluter00,lutsyshyn11} (space group No.~63, $Cmcm$),
$\rm Tb_2Ir_3Ga_9$ (TIG) contains alternating $\rm IrGa_2$ ($A$) and
$\rm Tb_2Ga_3$ ($B$) layers.  Along $\vc $, these layers stack to form an $A-B-A'-B'$ sequence,
where layers $A'$ and $B'$ result from a mirror-plane operation
on layers $A$ and $B$. The magnetic Tb atoms form a slightly distorted
honeycomb network, with two short Tb-Tb bonds of 4.28~\AA~along $\va $ and
four longer bonds of 4.38~\AA~rotated approximately $\pm 60^\circ$ away from $\va$ [Fig.~1(b)].

\begin{figure}[thb!]
 \includegraphics[width=3.3in]{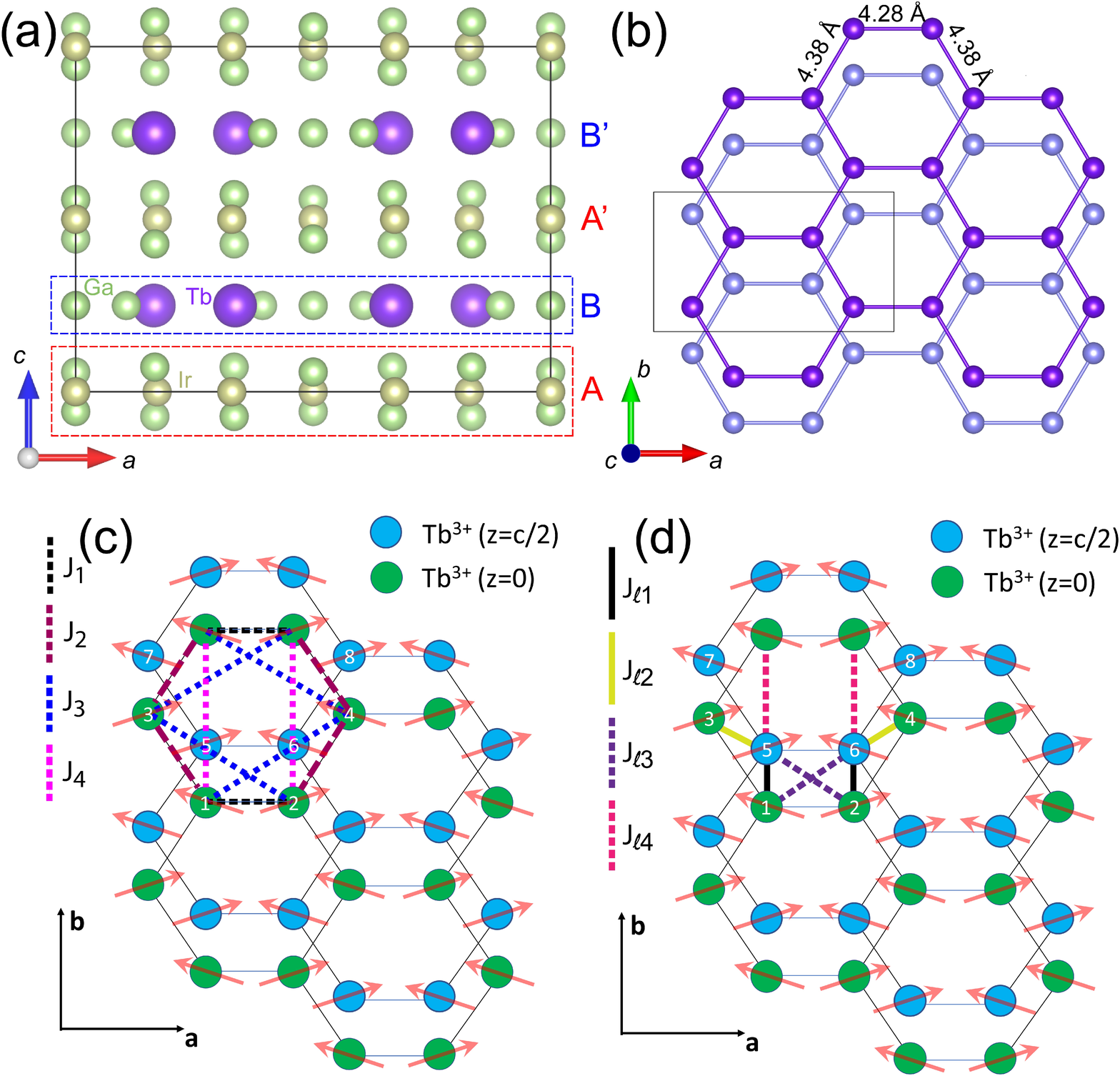}
    \caption{(a) The crystal structure of TIG projected onto the $ac$
    plane. The structure is composed of stacked $(AB)_2$ layers, where $A$ is a
    buckled $\rm IrGa_2$ layer (Ir atoms form a triangle lattice) and $B$ is a
    $\rm Tb_2Ga_3$ layer (Tb atoms form a pseudo-honeycomb lattice).
    (b) The network of Tb ions viewed from the $\vc$ axis.
    (c)-(d) The canted AFM spin configuration with magnetic space
    group $Cm'cm'$. $J_1$, $J_2$, $J_3$, and $J_4$ are the in-plane exchange interactions
    with Tb-Tb distances of 4.28, 4.38, 7.52, and 7.54 \AA; $J_{l1}$, $J_{l2}$, $J_{l3}$,
    and $J_{l4}$ are the inter-layer exchange interactions with Tb-Tb distances of
    5.36, 5.37, 6.86, and 6.89~\AA, all at room temperature.
    }
\end{figure}

The crystal field splits the $L=3$, $S=3$, and $J=6$ levels of Tb$^{3+}$ into
a low-lying non-Kramers doublet \cite{molavian07, curnoe13} and 11 higher levels.
Due to the interaction energies between ions, this non-Kramers doublet 
hybridizes with a higher-energy doublet to form a Kramers doublet.  So as in the other SL candidates,
the magnetic Tb$^{3+}$ moments can be treated as $J_{{\rm eff}}=1/2$ moments.

Hexagon-shaped single crystals of TIG, with typical size of a few
millimeters (mm) on the edge and 1-2~mm in thickness, were grown using a
Ga-flux method \cite{khan19b}.
The magnetization was measured using a Quantum Design SQUID.
Neutron diffraction was
performed on the HB1A triple axis spectrometer at the High Flux Isotope
Reactor (HFIR) and on the CORELLI and TOPAZ  diffractometers at
the Spallation Neutron Source, all at ORNL. Diffraction studies were made
on a naturally cleaved single crystal with dimensions $\rm 2\times2\times1~mm^3$.
Sample temperature $T$ was controlled using
the orange cryostat at HB1A, closed-cycle refrigerator (CCR) at CORELLI, and
Cryomech P415 pulse tube cryocooler at TOPAZ.

Inelastic neutron scattering (INS) studies were performed on the HB1 and HB3 triple axis
spectrometers at the HFIR.
A sample assembly of 35 single crystals (total mass $\sim$ 3.4~gram, mosaicity
$\sim1.5^\circ$) was aligned in the $(H, 0, L)$ scattering plane to probe
magnetic excitations in the basal plane and between layers.  Due to the weak
orthorhombic distortion, no attempt was made to align the pseudo-hexagonal crystals
along their common orthorhombic axis $\va$.  A CCR was used to
regulate the temperature for the INS measurements at HB1 and HB3.

The absence of a detectable signal from x-ray magnetic circular dichroism (XMCD)
measurements at the Ir $L$ edges places the upper limit for the Ir moments
at $0.01\,\mu_{\rm B}$\cite{khan19b}.
In the same work, the refined neutron powder diffraction pattern
indicated that the Tb spin configuration can best be described as
collinear order in the basal plane with easy axis along $\va $, consistent with
the magnetic space group (SG) $Cm'cm'$.  Although canted antiferromagnetic (AFM) order is
allowed by this SG, introducing a ferromagnetic (FM) component along $\vb $
did not improve the refinement.

Analysis of the magnetic properties is simplified by the confinement of
the magnetic  moments to the Tb sites \cite{khan19b}.  Under an applied field along $\va$,
the magnetization $M_a(H)$ shows step-like transitions at 2.5 and 6.5~T.
With increasing field along $\vc $, $M_c(H)$ exhibits linear response.
While the $\vb$-axis magnetization $M_b$ shows similar linear behavior, the
hysteresis loop below 1~T indicates the presence of a FM component.

The nearly Ising character of the Tb moments was demonstrated by measurements
of the critical fields $B_{c1}$ and $B_{c2}$ as the field is rotated by an angle $\phi$
away from the $\va$ axis within the $ab$ plane.  Both $B_{c1}(\phi )\cos \phi $
and $B_{c2}(\phi )\cos \phi $ are almost independent of angle $\phi $ up
to about $\pi /3$. Therefore, the component of the field along the $\va$ axis predominantly
controls the magnetic phase transitions \cite{khan19b}.  Similar results were found for the
Ising-like compounds TbNi$_2$Ge$_2$ [\onlinecite{budko99}] and Y$_{1-x}$Tb$_x$Ni$_2$Ge$_2$
[\onlinecite{wiener00}], where the Ni atoms are non-magnetic because the
Stoner criteria is not satisfied \cite{shigeoka92}.

\section{neutron diffraction results}

Although the two-dimensional (2D) spin Hamiltonian employed in an earlier
study\cite{khan19b}  captured the key characteristics of the
exchange interactions and described the metamagnetic transitions, the magnetic
order derived from neutron powder diffraction is clearly three dimensional (3D).
However, the sizable Dzyalloshinskii-Moriya (DM) interaction that produces the
FM moment along $\vb$  was not observed in neutron powder diffraction. To
reconcile this inconsistency, a comprehensive study of the static spin order
and magnetic dynamics  using single crystals was undertaken.

\begin{figure}[thb!]
     \includegraphics[width=3.4in]{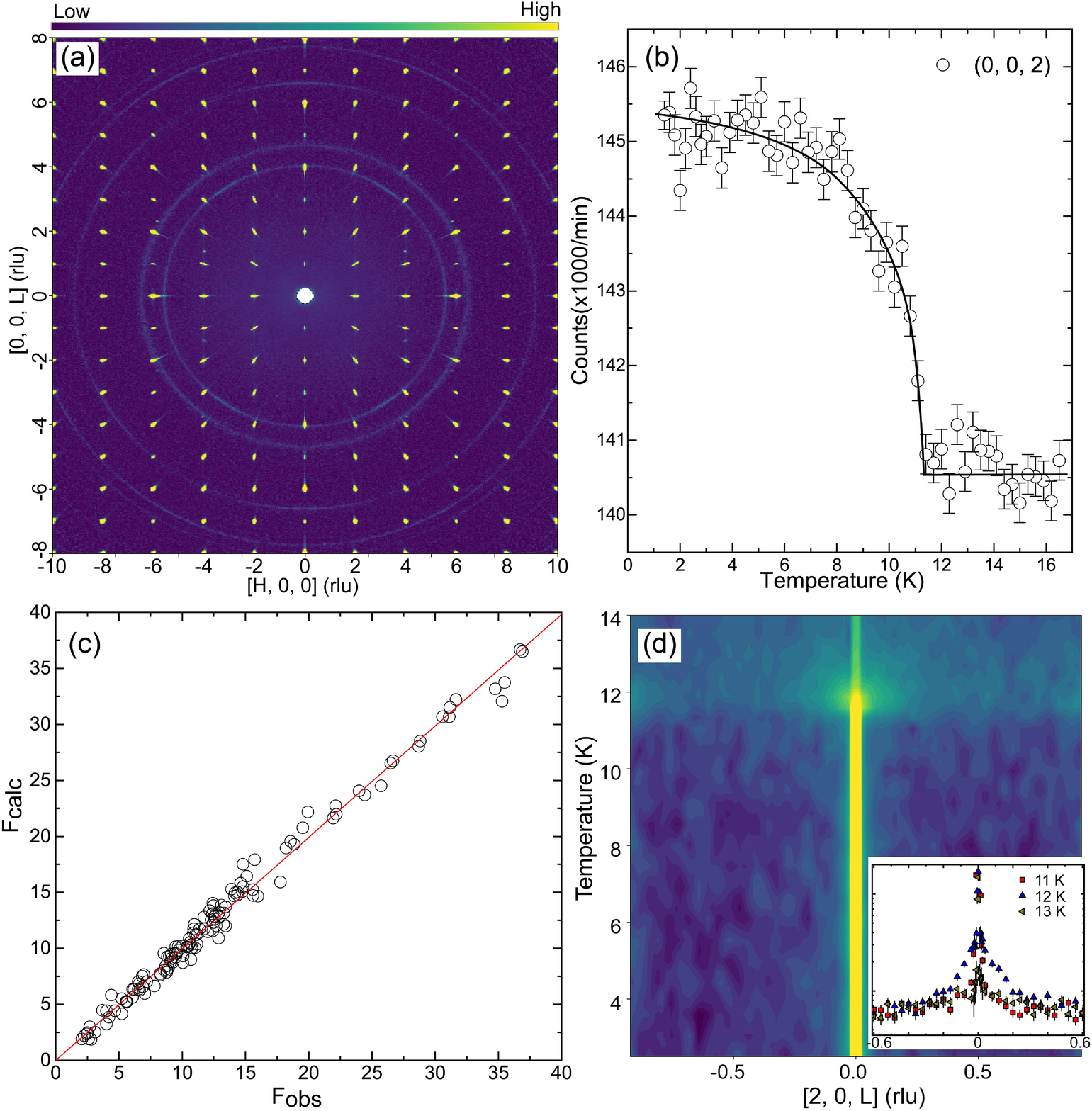}
     \caption{ (a) The contour plot of the neutron diffraction data in the
     $(H,0,L)$ scattering plane collected on CORELLI at $T=7$~K.
     (b) The $T$-dependence of the $(0,0,2)$ Bragg peak measured at HB1A.
     (c) Comparison between the observed and calculated structure factors,
     $\rm F_{obs}$ and $\rm F_{calc}$. The (red) line is a linear fit to the data points.
     (d) The $T$-dependence of the $L$-scan across the $(2,0,0)$ Bragg point.
     Inset shows the representative line-cut along the $[0,0,L]$ direction
     at 11, 12, and 13 K, with  prominent short range correlation at 12 K.
     }
\end{figure}

We first investigated the static magnetic order at low temperature.
Figure 2(a) provides a contour plot of the neutron diffraction data in the
$(H,0,L)$ scattering plane at 7~K measured
at CORELLI \cite{ye18c}. Consistent with neutron powder diffraction,
all observed reflections lie at integer indices, indicating that
the magnetic peaks coincide with the nuclear ones and have a propagation wavevector
$(0,0,0)$.  Group theory analysis indicates that the magnetic
representation $\rm \Gamma_{mag}$ for the magnetic Tb ion located at (0.336, 0.332, 1/4)
for SG $Cmcm$ can be decomposed as
$\rm \Gamma_{mag}=\Gamma_1+2\Gamma_2+\Gamma_3+2\Gamma_4+2\Gamma_5+\Gamma_6+2\Gamma_7+\Gamma_8$,
where $\Gamma_1$, $\Gamma_3$, $\Gamma_6$, and $\Gamma_8$ are one-dimensional (1D)
irreducible representations (IRs)
with moment only allowed along the $\vc$-axis,
while $\Gamma_2$, $\Gamma_4$, $\Gamma_5$, and $\Gamma_7$
are 2D IRs with moment permitted in both the $\va$ and $\vb$ directions
(the corresponding IRs and basis vectors are listed in Table 1).

Since the magnetization reveals a prevailing in-plane moment,
the 1D IRs with $\vc$-axis moment were not used to refine the
magnetic structure. The magnetic space groups (MSG) for the remaining
2D IRs are $Cm'c'm'$, $Cmcm'$, $Cm'cm'$, and $Cmc'm'$.
A mapping of the 3D reciprocal volume at the
CORELLI diffractometer at 7~K yields 393 reflections
that contain both magnetic and nuclear contributions.  Simultaneously
fitting both the crystal and magnetic structures reveals a
canted AFM structure best described by MSG $Cm'cm'$ [Fig.~2(c)].
In contrast to results of neutron powder diffraction,
this single-crystal study  identifies a small FM component along $\vb $.

Confirming this FM moment, Fig.\,2(b) plots the  thermal evolution of the
$(0,0,2)$ peak collected using a fixed incident energy at the
triple-axis spectrometer HB1A.  If canted order were absent, this purely
structural reflection would be $T$-independent.
Albeit weak, the abrupt enhancement (about 4\%) below $\rm T_N$ confirms the
FM component along $\vb $.  Summarized in Figs.~1(c)-1(d), the Tb moments form
a predominantly AFM state  along $\va$ canted by 6.7(3)$^\circ$
towards $\vb$.  At 7~K, the ordered moment along $\va $ is $\rm 17.8(4)~\mb$
per formula unit (f.u.), in  excellent agreement with magnetization measurements.

Since the Tb atoms form an orthorhombic rather than a true honeycomb lattice,
the collected single crystal diffraction data comprise three
unevenly populated structural and magnetic domains. The refinement on a single
piece of the crystal yields a domain volume fraction ratio of 11:77:12.
These three domains are described by rotation matrices: the first corresponds
to the crystal orientation matrix and the other two are given by
rotations of $\pm 60^\circ$ about $\vc $. The coexistence of those
twinned domains explains the strong magnetic Bragg reflections like
(2,0,$L=2n$).  An independent measurement at the TOPAZ diffractometer
on the same single crystal at 9.6~K (closer to the transition) confirms
the refinement results for the canted spin configuration.

\begin{table}[ht!]
\caption{Irreducible representation (IR), magnetic space group (MSG), and
Basis vectors (BVs) for the space group $Cmcm$ with magnetic propagation
vector ${\bf k}=(0,0,0)$. The Tb atoms of the nonprimitive basis are located at
$1:(x,y,z)$,~$2:(-x,-y,z+1/2)$,~$3:(-x,y,-z+1/2)$,~$4:(x,-y,-z)$,
with $x=0.336,~y=0.332,~z=1/4$. $\Gamma_1$,~$\Gamma_3$,~$\Gamma6$, and $\Gamma_8$
are 1D IRs with moments along the $\vc$-axis, $\Gamma_2$,~$\Gamma_4$,~
$\Gamma_5$, and $\Gamma_7$ are 2D IRs with moments allowed in the basal
plane.
}
\label{tab:BV}
\begin{ruledtabular}
\begin{tabular}{rrrrrr|rrrrrr}
IR/MSG & BV & No. & \multicolumn{3}{r}{component} & IR/MSG & BV & No. &\multicolumn{3}{r}{component} \\
           &          &   &$m_a$ &$m_b$ &$m_c$ & &          &  &$m_a$  &$m_b$ &$m_c$ \\
    \hline
$\Gamma_1$ & $\psi_1$ & 1 &  0 & 0& 1 & $\Gamma_3$ & $\psi_2$ &1 & 0 & 0& 1 \\
$Cmcm$     &          & 2 &  0 & 0& 1 & $Cm'c'm$   &          &2 & 0 & 0& 1 \\
           &          & 3 &  0 & 0&-1 &            &          &3 & 0 & 0& 1 \\
           &          & 4 &  0 & 0&-1 &            &          &4 & 0 & 0& 1 \\
$\Gamma_6$ & $\psi_3$ & 1 &  0 & 0& 1 & $\Gamma_8$ & $\psi_4$ &1 & 0 & 0& 1 \\
$Cmc'm$    &          & 2 &  0 & 0&-1 & $Cm'cm$    &          &2 & 0 & 0&-1 \\
           &          & 3 &  0 & 0&-1 &            &          &3 & 0 & 0& 1 \\
           &          & 4 &  0 & 0& 1 &            &          &4 & 0 & 0&-1 \\
$\Gamma_2$ & $\psi_5$ & 1 &  1 & 0& 0 &$\Gamma_4$ & $\psi_7$ & 1 & 1 & 0& 0 \\
$Cm'c'm'$  &          & 2 & -1 & 0& 0 & $Cmcm'$    &          & 2 &-1 & 0& 0 \\
           &          & 3 & -1 & 0& 0 &           &          & 3 & 1 & 0& 0 \\
           &          & 4 &  1 & 0& 0 &           &          & 4 &-1 & 0& 0 \\
           & $\psi_6$ & 1 & 0 & 1& 0 &            & $\psi_8$ & 1 & 0 & 1& 0 \\
           &          & 2 & 0 &-1& 0 &            &          & 2 & 0 &-1& 0 \\
           &          & 3 & 0 & 1& 0 &            &          & 3 & 0 &-1& 0 \\
           &          & 4 & 0 &-1& 0 &            &          & 4 & 0 & 1& 0 \\
$\Gamma_5$ & $\psi_9$ & 1 & 1 & 0& 0 &  $\Gamma_7$ & $\psi_{11}$ &1 & 1 & 0& 0 \\
$Cm'cm'$   &          & 2 & 1 & 0& 0 &  $Cmc'm'$   &          &2 & 1 & 0& 0 \\
           &          & 3 &-1 & 0& 0 &             &          &3 & 1 & 0& 0 \\
           &          & 4 &-1 & 0& 0 &             &          &4 & 1 & 0& 0 \\
           &$\psi_{10}$ & 1 & 0 & 1& 0 &           & $\psi_{12}$ &1 & 0 & 1& 0 \\
           &          & 2 & 0 & 1& 0 &             &          &2 & 0 & 1& 0 \\
           &          & 3 & 0 & 1& 0 &             &          &3 & 0 &-1& 0 \\
           &          & 4 & 0 & 1& 0 &             &          &4 & 0 &-1& 0 \\
  \end{tabular}
\end{ruledtabular}
\end{table}

\begin{figure}[thb!]
     \includegraphics[width=3.3in]{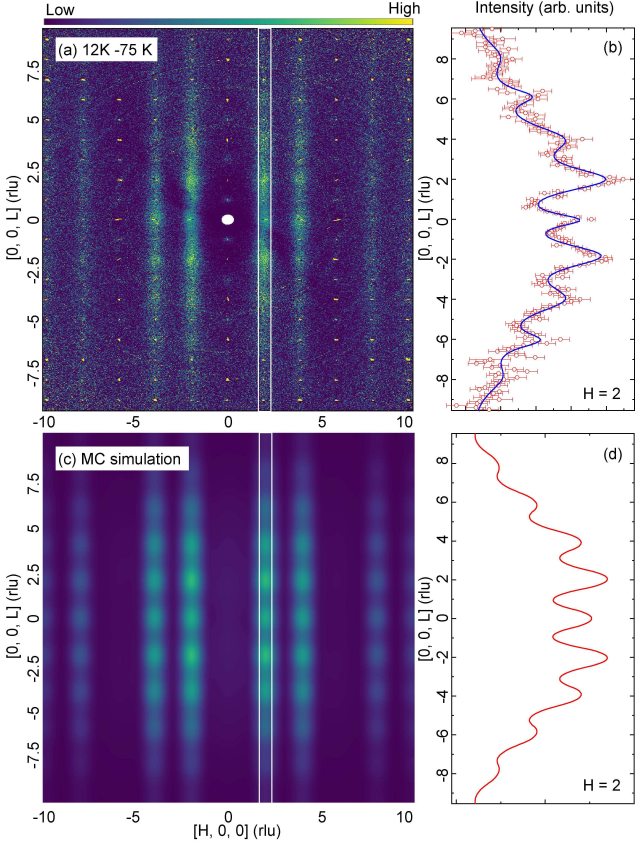}
     \caption{ (a) Contour plot of the neutron diffraction data in the
     $(H,0,L)$ scattering plane. A rod-like feature along $[0,0,L]$
     indicates the short-range magnetic correlations along the $\vc$
     axis. (b) For the line cut along $[2, 0, L]$, Lorentizan
     profiles appear at even indices. (c) Monte-Carlo simulation of the
     magnetic diffuse scattering just above the transition
     using the magnetic exchange parameters
     in Tab.~II. (d) The corresponding line cut along $[2,0,L]$.
     }
\end{figure}

Notably, the single-crystal study indicates significant magnetic
correlation between honeycomb layers just above $T_N$, as shown by the
$T$-dependence of the $L$-scan across $(2,0,L)$ [Fig.~2(d)]. A more detailed
characterization of the spin-spin correlation is given by the $(H,0,L)$
slice in Fig.~3(a), which shows the $T=12.5$~K data after
the $75$~K data is subtracted as background.  Short-range spin fluctuations along
$[0,0,L]$ are prominent at $H=-8,-4,-2,2,4,$ and 8.
The 1D line cut at $H=2$ with $\Delta H=\pm0.2$ shown in Fig.~3(b)
can be fit as the summation of multiple Lorentzian profiles peaked at $L=2n$
on top of a broad Lorentzian background. The half-width/half-maximum (HWHM) of these profiles
ranges from 0.60 to 0.98 reciprocal lattice unit (rlu), corresponding to a magnetic
correlation length from $9.6$ to $15.8$~\AA, which is longer than the nearest
neighbor Tb-Tb distances ($\sim 5.4$~\AA) between honeycomb layers.  Whereas the
magnetic diffuse scattering in pure 2D systems should exhibit featureless
fluctuations between the layers, the observed multiple peaks
indicate considerable 3D magnetic correlations along $\vc $ and are
consistent with the spin dynamics analysis presented below.

\begin{figure*}[ht!]
     \includegraphics[width=5.5in]{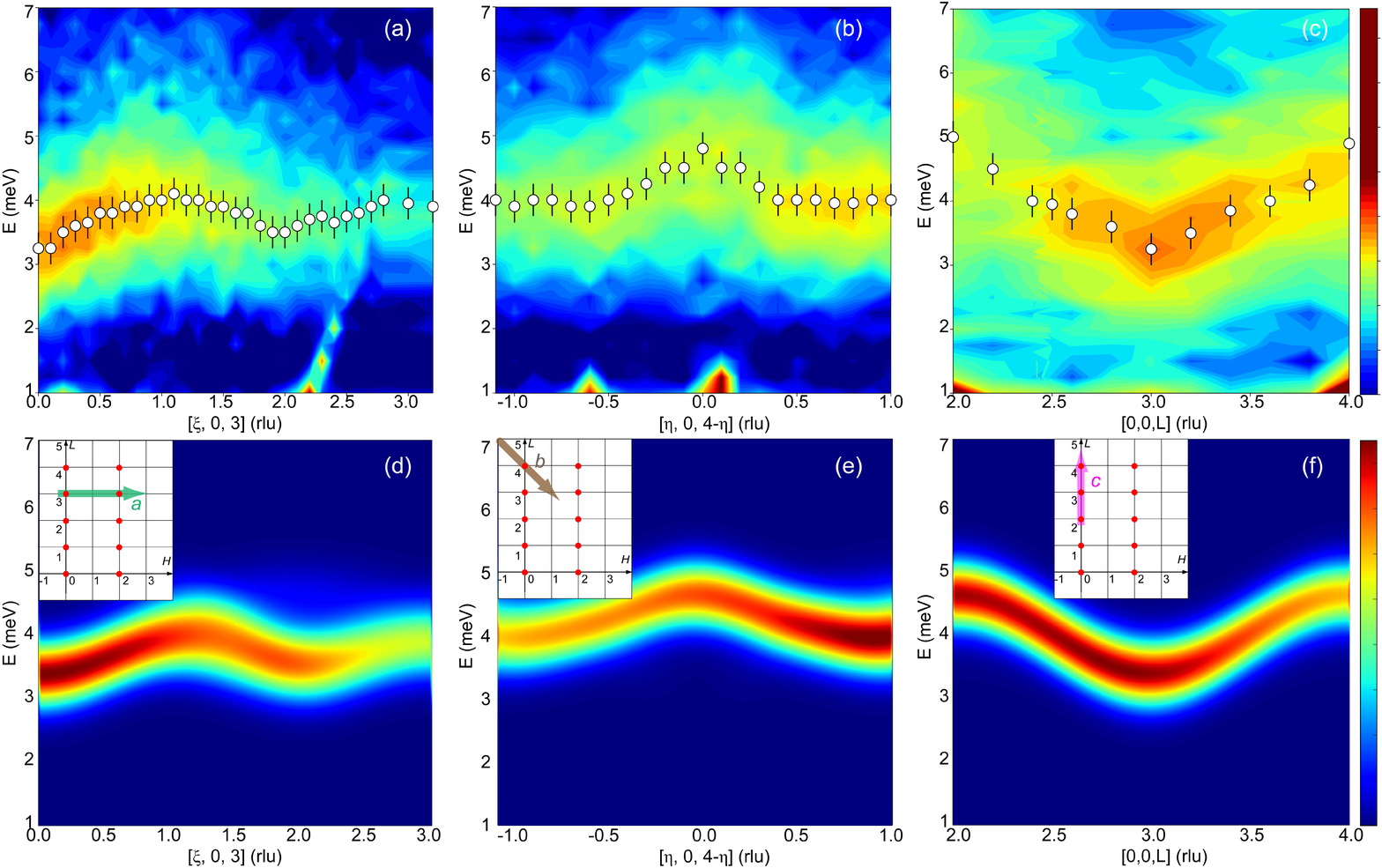}
     \caption{SW dispersion spectra of $\rm Tb_2Ir_3Ga_9$ along (a) $[\xi,
     0, 3]$, (b) $[\eta, 0, 4-\eta]$, and (c) $[0, 0, L]$. The
     corresponding calculated spectra are shown in panels (d)-(f). Insets show the
     schematics of the scan directions in the $(H,0,L)$ scattering plane.
     The dispersion-like feature for energy transfer below 3 meV in
     panel (a) arises from the tail of the resolution function sweeping
     through the neighboring Bragg peak in the focusing geometry.
     }
\end{figure*}

\section{inelastic neutron scattering study}

An earlier description \cite{khan19b} of TIG was based on a model with
anisotropic exchange along the bond direction $\vR_i -\vR_j$ between Tb$^{3+}$ ions
in each layer. That model provided an excellent description of the magnetization data.
However, as discussed further below, it does not provide an adequate
description of the spin dynamics.  Therefore,
we now study TIG using an $xxz$ model, which has been previously used
to describe other layered honeycomb systems  \cite{Maksimov16, Nair18, Matsumoto20}
and has also been proposed for rare-earth compounds \cite{Mackintosh72}.
The Hamiltonian is given by
\begin{eqnarray}
{\cal H}&=&-\frac{1}{2}\sum_{i, j}J_{ij}^{xy} \bigl\{ S_{ix} S_{jx}+ S_{iy}S_{jy}\bigr\}   -\frac{1}{2}\sum_{i, j} J_{ij}^z \, S_{iz}  \, S_{jz}    \nonumber \\
&-&\frac{1}{2}\sum_{i, j}J_{l\, ij}^{xy} \bigl\{ S_{ix} S_{jx}+ S_{iy}S_{jy}\bigr\}   -\frac{1}{2}\sum_{i, j} J_{l\, ij}^z \, S_{iz}  \, S_{jz}    \nonumber \\
&-& K_x \sum_i {S_{ix}}^2  -K_z \sum_i {S_{iz}}^2
\nonumber \\
&-&\frac{1}{2}\sum_{i, j}^{1st, \, 2nd} {\bf D}_{ij} \cdot (\vS_i \times \vS_j )- \mb \sum_{i, \alpha } g_{\alpha \alpha} B_{\alpha }S_{i\alpha },
\label{ham}
\end{eqnarray}
which replaces the total angular momentum $\bf J_i$ of Tb$^{3+}$ by an effective spin $\bf S_i$ at site $i$.
Exchanges $J_n$ act between spins within each $ab$ plane and exchanges $J_{ln}$ act
between spins on neighboring planes [Figs.~1(c)-(d)].  Each exchange
interaction contains an $xy$ part $J_{ij}^{xy}$
that couples the $x$ and $y$ components of the spin and a $z$ part $J_{ij}^z$
that couples the $z$ spin components.

Although single-ion anisotropy is expected to vanish within the $J_{{\rm eff}}=1/2$,
``pseudo"-doublet state of Tb$^{3+}$ [\onlinecite{molavian07, curnoe13}],
easy-plane and easy-axis anisotropies $K_z$ and $K_x$
confine the spins in the basal plane and align them along $\va$.  These single-ion
anisotropy terms will be further discussed in the conclusion.

While the nominal $g$-factor for $S=L=3$ and $J=6$ moments is $g=3/2$, we treat
the diagonal components $g_{xx}$, $g_{yy}$, and $g_{zz}$ of the $g$-tensor as fitting parameters.
Initial fitting results indicated that the  nearest-neighbor interactions (both $xy$ and $z$ components) $J_1$
and $J_{l1}$ can be set to zero.
The $z$ components of $J_3$, $J_4$, and $J_{l4}$ are small and neglected.
It is permissible to take $J_3^{xy}=J_4^{xy}$, which is expected from the
nearly identical distances 7.52 and 7.54\,\AA\,spanned by those interactions.

The DM interaction ${\bf D}_{ij}=D\vc $ is allowed by the broken inversion
symmetry caused by the alternation of the Ir$^{4+}$ ions on either side of the
Tb-Tb bond moving around a hexagon in the honeycomb lattice.
This DM interaction couples both nearest-neighbor spins 1 and 2 or 3 and 4 separated by
4.28~\AA, and next-nearest neighbor spins 1 and 3 or 2 and 4 separated by 4.38~\AA.
Whereas $D$ cants the spins away from the $\va $ axis, the exchange interactions and the easy-axis
anisotropy $K_x$ favor a collinear state.  Minimizing the total energy, the canting angle
is given by
\begin{equation}
\label{th}
\theta=\frac{1}{2}\tan^{-1}\Biggl\{ \frac{3D}{J_1+2J_2+2J_{l1}+2J_{l4}-K_x}\Biggr\}.
\end{equation}
Since $M_0= 2g_{yy}\mb S \sin \theta \approx 1.22\,\mb$ is the canted moment/f.u. along $\vb$ observed by
magnetization measurements [Fig.~5(b)], Eq.(\ref{th}) fixes $D$ in terms of the other
model parameters and $M_0$.   Hence, the total of fitting parameters is 13 [Table II].

Even though neutron diffraction measurement on one single crystal revealed
an uneven  distribution of domains, we made no effort to align the
orthorhombic axes of the 35 small crystals.  Due to the large number of single
crystals, we expect an equal fraction of those
crystals to have their orthorhombic axes along $(1,0,0)$, $(1/2,\sqrt{3}/2,0)$,
and $(-1/2, \sqrt{3}/2,0)$ for domains 1, 2, and 3, respectively.  This is confirmed
by least square fit of the corresponding domain contributions to the magnetic peak intensities.
For scans along $(H,K,L)$ with $K=0$, domains 2 and 3 have the same set
of SW branches but domain 1 has a different set.

The SW dynamics at zero field is evaluated by  taking sites 1 and 4 (5 and 8)
and sites 2 and 3 (6 and 7) on layer 1 (2) to be identical.  Since the magnetic
unit cell contains 4 distinct spins, each domain produces 4 SW modes.
For scans  along $(\xi ,0, 3)$ and $(\eta ,0,4-\eta )$, our model predicts
8 SW branches.  For the scan along $(0,0,L)$, each domain produces the same spectra
and our model predicts 4 SW branches.

However, Figs.~4(a-c) reveal a single wide SW branch for each scan.
To compare the calculated and measured SW frequencies,
we perform a weighted average over the calculated frequencies at each wavevector:
\begin{equation}
\omega_{\rm av}(\vq ) =  \frac{ \sum_n \omega_n(\vq ) S_n(\vq ) }{\sum_n S_n(\vq )},
\end{equation}
where the weight $S_n(\vq )$ is obtained
from the spin-spin correlation function
$S_{\alpha \beta }(\vq ,\omega )$ using
\begin{eqnarray}
S(\vq ,\omega ) &=& \Bigl\{ \delta_{\alpha \beta } -\frac{q_{\alpha }q_{\beta }}{q^2} \Bigr\} S_{\alpha \beta } (\vq , \omega ) \nonumber \\
&=& \sum_n S_n(\vq )\, \delta (\omega - \omega_n(\vq )).
\end{eqnarray}
To order $1/S$ in the Holstein-Primakoff expansion \cite{SWBook2018},
each mode produces a delta function $\delta (\omega - \omega_n(\vq ))$ with weight $S_n(\vq )$.

Our original fits based solely on the weighted SW frequencies produced a
wide spread in SW intensities that was inconsistent with the measurements.
Therefore, we constrained the observed spread in frequencies to be greater
than or equal to the calculated spread $2\Delta \omega (\vq )$, where
\begin{equation}
\Delta \omega (\vq )^2 = \frac{ \sum_n (\omega_n (\vq )-\omega_{\rm av}(\vq ))^2 S_n(\vq )}{\sum_n S_n(\vq )}.
\end{equation}
The cost function in $\chi^2_{\rm INS}$ used an experimental uncertainty in the peak
SW frequencies of $\sigma_{\omega } = 0.25$\,meV for both instruments HB1 and HB3.

To evaluate the magnetic $\chi^2_{\rm mag}$, we used an experimental
uncertainty in the magnetization of $\pm 6\%$ for field above $B_{c1}$
along $\va $ and for all fields along $\vb $ and $\vc $.  The
calculated critical fields $B_{c1}$ and $B_{c2}$ along $\va $ were constrained to agree
with the measured critical fields.
In addition, $B_{cn}(\phi )\cos \phi $ was constrained to be nearly
independent of the angle $\phi $ between the applied field and the $\va$ axis
within the $ab$ plane up to $\phi = \pi /3$. The 13 fitting  parameters were
then determined by minimizing $\chi^2 = \chi^2_{{\rm mag}} +\chi^2_{{\rm INS}}$.

\begin{table}[ht!]
    \caption{The in-plane and out-of-plane exchange interaction parameters $J_i$ and $J_{li}$,
    easy-axis and easy-plane anisotropies $K_x$ and $K_z$ and DM exchange interaction $D$, units in meV.
    Values in parentheses are the error bars.}
\begin{ruledtabular}
\centering
\begin{tabular}{ccc}
    \multicolumn{1}{c}{parameter}   & \multicolumn{2}{c}{value} \\
\hline
    & $xy$ & $z$\\
    $J_1$ & $0$   & $0$ \\
    $J_2$ & $-0.016(2)$    & $-0.05(2)$ \\
    $J_3=J_4$ &  $0.007(1)$   & 0   \\
    $J_{l1}$ & $0$     & $0$ \\
    $J_{l2}$ & $-0.027(2)$ & $0.09(2)$ \\
    $J_{l3}$ & $0.014(6)$  & $-0.16(6)$ \\
    $J_{l4}$ & $-0.014(2)$ & 0 \\
    $K_x$ & \multicolumn{2}{c}{$0.09(1)$}  \\
    $K_z$ & \multicolumn{2}{c}{$-0.84(6)$}  \\
    $D$   & \multicolumn{2}{c}{$-0.0066$}  \\
    $g_{xx}$ & \multicolumn{2}{c}{$1.38(1)$} \\
    $g_{yy}$ & \multicolumn{2}{c}{$1.51(3)$} \\
    $g_{zz}$ & \multicolumn{2}{c}{$1.59(6)$} \\
\end{tabular}
\end{ruledtabular}
\end{table}

\section{Fitting results}

To compute the spectra, the delta-function intensities
$S_n(\vq )\delta (\omega -\omega_n(\vq))$ were convoluted over a Lorentzian with
width $\nu = 0.5$\,meV, which is close to the instrumental resolution for both
HB1 and HB3, and then multiplied by the square of the magnetic form factor
$f(q)$ for Tb$^{3+}$.  Results for the calculated magnetization
and inelastic intensities are plotted in Figs.\,4(d)-(f) and in Figs.\,5(a)-(c).
These results are quite
satisfactory with a few reservations.  First, the calculated
intensity along $[\xi ,0, 3]$ is fairly large up to $\xi =3$ while the observed
intensity drops off rapidly above $\xi = 2$.  Second, the calculated intensity
along $[0,0,L]$ peaks to the left of $L=3$ while the observed intensity peaks to the
right.  Third, the calculated magnetization is slightly too small for fields along
$\vb $ and $\vc $ in Figs.\,5(b) and (c).  By contrast, the calculated magnetization
for field along $\va $ in Fig.\,5(a) is slightly too large in the plateau
between 2.5 and 6.5 T.

\begin{figure}[thb!]
     \includegraphics[width=3.2in]{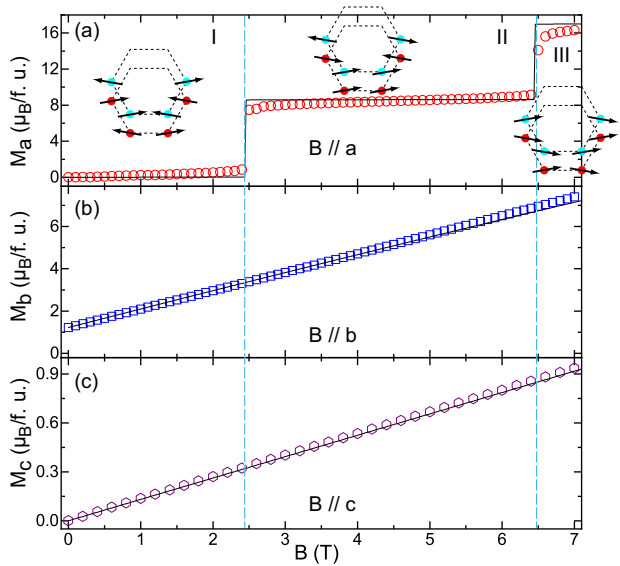}
     \caption{Magnetization $M(H)$ with field applied along the three crystallographic
     axes up to 7~T at 1.8~K. Open symbols are experimental data;
     solid lines are the best fits described in the text.
     The spin configurations in three distinct region
     with field $B\parallel \va$ are sketched in panel (a).
     }
\end{figure}

The microscopic parameters that minimize the total $\chi^2$ are given in Table II.
The resulting $\underline{g}$-tensor parameters have an average value
$g_{{\rm av}}=(g_{xx}+g_{yy}+g_{zz})/3$ of 1.50(3), overlapping with the nominal 
$J=6$ value of $g=1.5$. This result is consistent with measurements \cite{khan19b} 
of the Curie-Weiss susceptibility, which gives an effective moment of 
10.3\,$\mb $/Tb, close to the free ion value of 9.7\,$\mb $/Tb when $g=1.5$.

By far the largest energy among the fitting parameters is the easy-plane anisotropy $K_z \approx -0.83$\,meV.
A rough estimate for $K_z$ can be obtained from the observed  magnetization when
a field is applied along $\vc$.  Neglecting the exchange interactions, the energy per
spin is given by
\begin{equation}
E = K_z S^2 \cos^2 \theta - \mb g HS \sin\theta ,
\end{equation}
where $\theta $ is the canting angle of the spin towards $\vc $.  Minimizing this energy
with respect to $\theta $ gives a magnetization per f.u. of
\begin{equation}
M_z = 2g_{zz}\mb S \sin \theta = \frac{\mb^2 g^2 H }{\vert K_z\vert }.
\end{equation}
Using the experimental result $M_z = 0.94\,\mb$/f.u. at 7\,T (corresponding to a
tilt angle of $\theta = 3.1^\circ$) and the value $g_{zz}$ = 1.59 from Table II, we find
$K_z \approx -0.98$\,meV (a value of $-0.88$\,meV was found in Ref.\,[\onlinecite{khan19b}]).
Thus, a large value of $K_z$ is required to explain the
small magnetization when a field is applied along $\vc $.  For bulk Tb in a hexagonal
close-packed structure, Rhyne {\em et al.}~\cite{Rhyne68} reported a tilt angle
of $8.6^\circ $ in a 7\,T field, corresponding to $K_z \approx -0.32\,$meV, less than half
the size of the one reported here. A similar analysis based on the change in magnetization
of $6.2\,\mb$/f.u.~in a 7\,T field along $\vb $ yields $K_x \approx 0.13$\,meV, which is
larger than our fitting result $0.09$\,meV because the $xy$ exchange energy also
strongly favors an AFM state.

Another remarkable feature of these results is that the $z$ exchange couplings
are substantially larger than the $xy$ couplings. To gain further insight,
we minimized $\chi^2_{{\rm mag}}$ without any dynamical contribution with respect
to five $xy$ exchange parameters and two anisotropies:
$J_2^{xy}=-0.014$, $J_3^{xy}=0.007$, $J_{l2}^{xy}=-0.027$,
$J_{l3}^{xy}=0.014$, $J_{l4}^{xy}=-0.016$, $K_x=0.08$, and $K_z=-0.84$, all in meV.
This static fit also gives $g_{xx}=1.39$, and $g_{yy}=g_{zz}=1.49$.  While $\chi^2_{{\rm mag}}$ slightly
decreases from 0.30 for the $xxz$ model with dynamical input to 0.29 for the $xx$
model without dynamical input, the resulting $xy$ exchange parameters
are close to those obtained in Table II from fitting the full $\chi^2
=\chi^2_{{\rm mag}}+\chi^2_{{\rm INS}}$.  Hence, the $z$ exchange couplings are {\it not}
required to explain the magnetization measurements.

The earlier model in Ref.\,[\onlinecite{khan19b}] used eight parameters to
explain the magnetization, fixing $g=1.5$ but adding hexagonal anisotropy.
By comparison, the model described above uses
ten parameters, including $g_{\alpha \alpha }$ but neglecting hexagonal anisotropy. In both models,
the exchange between spins 1 and 3 or 2 and 4 along a side of the hexagon is greater
than the exchange between spins 1 and 2 or 3 and 4 along the top or bottom of the hexagon.
Hence, bond-dependent exchange is required to understand the magnetization
measurements of TIG.

Assuming now that the exchange interactions are isotropic ($J_{ij}^{xy}=J_{ij}^z$),
minimizing the total $\chi^2$ with respect to all eight exchange parameters
gives $\chi^2=0.65$, which is greater than the value $0.33$ obtained using the
{\it anisotropic} parameters in Table II.  Hence, the five large $z$ exchange components
in Table II are required to explain the inelastic measurements.
Using a fitting technique that constrains the frequency width of the inelastic spectra,
we believe that our model contains the minimum number of  parameters that can adequately describe TIG.

The fitting result $g_{xx}=1.38(1)$ gives the saturation magnetization
$16.5(2)\,\mb $/f.u. and the ordered moment of $8.3(1)\,\mb $ for field along $\va $.
For comparison, the ordered moment $8.9(2)\,\mb $ obtained from neutron diffraction
measurements gives $g=1.48(3)$.

As an additional check on our results, we compare the observed \cite{khan19b}
transition temperature of 12.5\,K with the mean-field (MF) N\'eel temperature
evaluated for Ising spins:
\begin{equation}
T_{{\rm N }}^{{\rm MF}} =z\vert J_{xy}\vert  \frac{S(S+1)}{3},
\end{equation}
where
\begin{eqnarray}
zJ_{xy} &=& 2J_2^{xy}-4J_3^{xy}-2J_4^{xy}
+2(J_{l1}^{xy}-J_{l2}^{xy}\nonumber \\
&-&J_{l3}^{xy}+J_{l4}^{xy}).
\end{eqnarray}
Since $zJ_{xy} \approx -0.074$\,meV,
$T_{{\rm N}}^{{\rm MF}}=$\,12.0\,K
is close to the observed transition temperature of 12.5\,K.

Finally, the $xxz$ spin Hamiltonian and the corresponding exchange parameters are
checked by calculating the diffuse scattering  near the transition.  A magnetic
super cell is constructed containing $8\times8\times8$ chemical unit cells
with 4096 Tb ions (8 atoms per chemical unit cell).  Using the parameter values
in Tab.~II, a forward cluster Monte-Carlo simulation \cite{donoriodemeo92} is
performed just above the transition temperature of $T=12.5$~K starting with the initial
ground state configuration.
After 1000 Monte-Carlo cycles (on average, one cycle
visits each of the 4096 atoms once), the diffuse scattering pattern is
calculated including the contributions of each of the three domains.

The resulting diffuse scattering pattern reveals significant 3D spin correlations.
Figs.~3(c)-(d) show the calculated diffraction pattern in
the $(H,0,L)$ plane and the line cut along the $[0,0,L]$.
The agreement between experiment and theory is excellent:
strong streak-like diffuse scattering appears at $H=2,4,8$ but is weak at $H=6$ and
the profile along $(0,0,L)$ has the same intensity distribution as in the experiment.
The peaks that appear at even $L$ are caused by the $\pm60^\circ$
domains while the peaks at odd $L$ are caused by the $0^\circ $ domain.
Remarkably, the  Monte-Carlo simulation gives the correct ground state
up to the transition temperature.

Monte-Carlo simulations also indicate that competing ground states lie close in
energy to the state in Figs.\,1(c) and (d) due to the sizeable AF
exchange $J_{l2}^{xy}\approx -0.027$\,meV between parallel spins.
This suggests that doping or pressure might produce a complex phase diagram.

\section{First-Principles Calculations}

To connect TIG's rather complex physical structure to its observed
magnetism, we performed first-principles calculations using the linearized
augmented plane-wave density functional theory code WIEN2K \cite{blaha2001wien2k}.
We  employed two standard approximations:  the generalized gradient approximation (GGA)
and the correlated version of this approach known as GGA+$U$, in which a Hubbard
$U$ (here chosen as 6\,eV) is applied to the Tb 4$f$ orbitals.  To account
for potential magnetoelastic effects
\cite{pokharel2018negative, sanjeewa2020evidence, yan2020type, chen2019suppression},
the experimental structure \cite{grin1989phases} of similar compounds
was optimized within the GGA in an assumed FM Tb configuration.
Muffin-tin radii of 2.17, 2.4 and 2.5 Bohr were chosen, respectively, for the Ga,
Ir and Tb atoms. Corresponding to the product of the smallest muffin-tin radius
and the largest plane-wave expansion wavevector, RK$_{max}$ was set to 8.0.
Given the rather detailed exposition in the previous work \cite{khan19b}, we have focused
on the interlayer exchange couplings.

Four magnetic states were studied - the previously mentioned FM configuration and
three AFM configurations. AF$_1$ has the 3 Tb-Tb planar  neighbors anti-aligned
and the next-nearest and next-next-nearest neighbor  planes FM and AFM coupled,
respectively; AF$_2$ has the same  planar orientation but next-nearest
and next-next-nearest planes AFM and FM coupled; and AF$_3$ is an interlayer
AF state with planar neighbors aligned  and next-nearest neighbor Tb planes antialigned.
In all cases, the same distorted honeycomb structure with lattice parameters
taken from experiment was assumed. The possible ground states given above correspond
to a substantially simplified set of configurations compared with the canted state
obtained from the neutron diffraction results, which is closest to AF$_2$.
Nevertheless, it captures important aspects of the relevant physics.

For simplicity, our calculations do not include SOC and so neglect the Tb
orbital moments.  Using GGA+$U$, all magnetic states have a substantial Tb
spin moment of 6.06\,$\mu_B$, slightly larger than the
spin moment of 5.83\,$\mu_B$ obtained using the straight
GGA and in good agreement with previous work \cite{khan19b}.
Within the GGA+$U$, AF$_1$ has the lowest energy, AF$_2$ and AF$_3$ lie 13 and 16\,meV
per Tb higher, respectively, and the FM state lies 43\,meV per Tb higher.

These energy differences were mapped onto a simple Heisenberg model including one
intralayer nearest-neighbor coupling $J^{(1)}$ and two next-nearest-neighbor
and next-next-nearest-neighbor interlayer couplings $J^{(2)}$ and $J^{(3)}$.
Using $S(S+1)=42$, we find $J^{(1)}=-0.22$\,meV, $J^{(2)}=-0.02$\,meV,
and $J^{(3)}= -0.18$~meV  - all AFM.  Notice that the next-next-nearest-neighbor
coupling $J^{(3)}$ is {\it not} substantially smaller than the nearest-neighbor
coupling $J^{(1)}$ despite the larger distances  spanned by $J^{(3)}$ (5.37\,\AA)
relative to the distances spanned by the  $J^{(1)}$ interactions (4.28 and 4.38\,\AA).
One may directly compare the result for $J^{(1)}$ to that for $J_2^{xy}$ and results
for $J^{(2)}$ and $J^{(3)}$ to those for $J_{l1}^{xy}$ and $J_{l2}^{xy}$ in Table II.

Although the distances 5.36 and 5.37~\AA~spanned by $J^{(2)}$ and $J^{(3)}$ differ
by just 0.01~\AA, those interactions are substantially different within GGA+$U$.
In agreement with the GGA$+U$ calculation, INS fits find that
$\vert J_{l1}^{xy}\vert \ll \vert J_{l2}^{xy}\vert$.
We ascribe the different magnitudes of those exchange couplings obtained from
GGA$+U$ and INS to the well-known
difficulties experienced by density-functional theory in quantitatively
describing 4$f$ physics. In any case, we reproduce both the right general
size of these interactions and their surprising, yet experimentally validated,
slow fall-off with distance.

\begin{figure}[thb!]
\includegraphics[width=3.2in,angle=0]{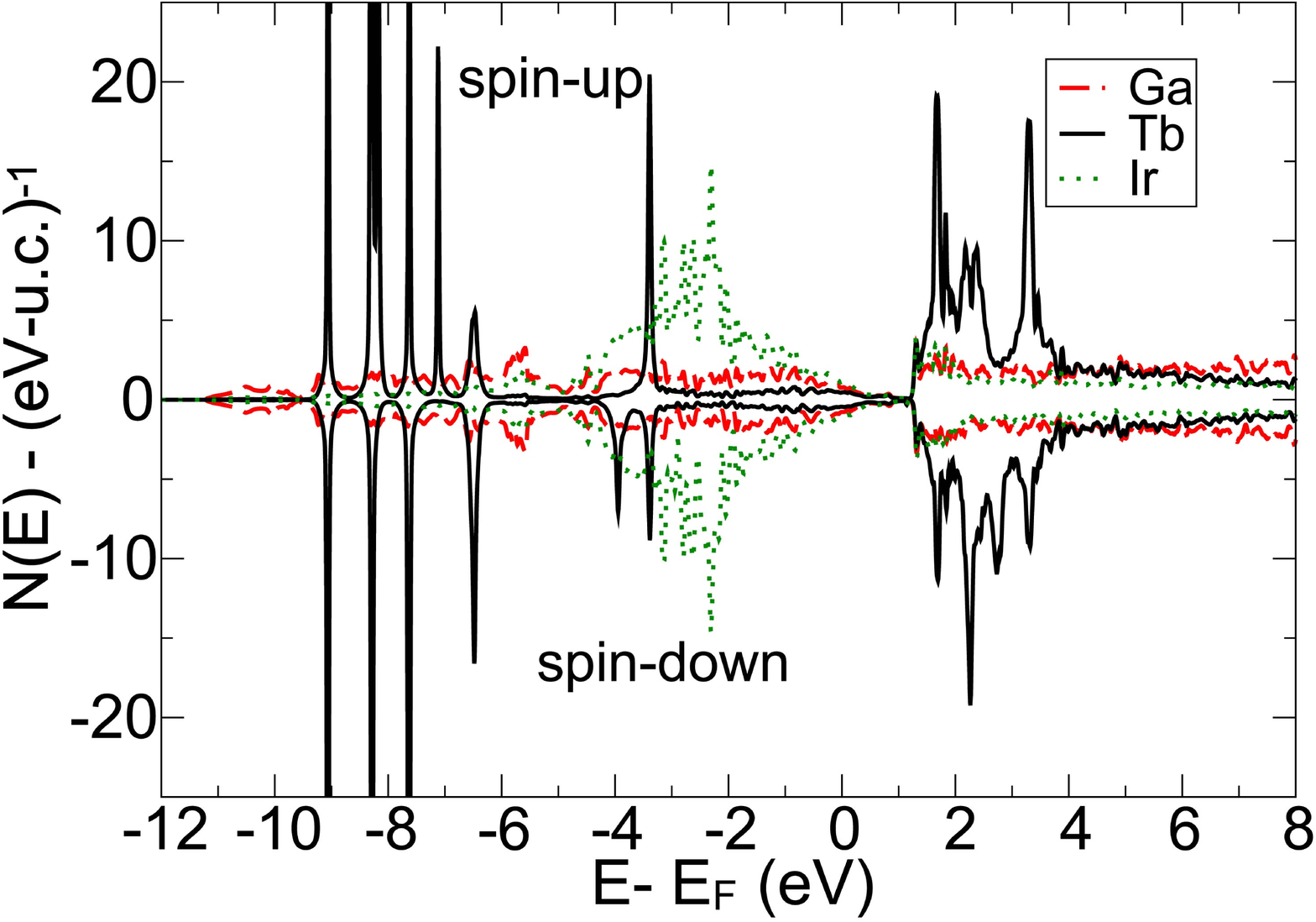}
\caption{The calculated density-of-states of Tb$_{2}$Ir$_{3}$Ga$_{9}$ in the
$\rm AF_1$ phase.}
\end{figure}

The calculated ground state density-of-states of TIG in Fig.\,6 indicates
the highly localized character of the Tb 4$f$ states along with the more
delocalized character of Ir and Ga. As in previous work \cite{khan19b},
the density-of-states is relatively low at the Fermi level, displays a
weak gap just above, and then exhibits peaks associated with the unoccupied Tb 4$f$
orbitals. The GGA+$U$ approach properly displaces the Tb 4$f$ states above and
below the Fermi level.

With multiple Ga atoms between the Tb planes, TIG contains several possible
indirect exchange or super-exchange pathways.
Indeed, recent work \cite{williams2016extended,sirica2020nature}
for 3$d$ compounds finds that such pathways can produce large exchange interactions
even at distances substantially exceeding 5\,\AA. Despite the typical
localization of 4$f$ moments, it is possible that the combination of Tb and
Ga produces a similar effect.

\section{Discussion and Conclusion}

It is well-known that the charge distribution of the Tb$^{3+}$ $f$-orbital is highly
anisotropic \cite{Rhyne72}.  In our numerical fits to the inelastic spectra,
both first-neighbor interactions $J_1$ and $J_{l1}$ within each layer or
between layers are negligible.  While $J_1$ couples sites with
$\vR = \vR_i -\vR_j$ along $\va $ ($R=4.28$\,\AA), $J_{l1}$ couples sites with
$\vR \cdot \va =0$ ($R=5.33$\,\AA). This suggests that for small $R$, the exchange
couplings satisfy
\begin{equation}
J_{ij} \approx \frac{J(R)}{R^4} \Bigl\{ (\vR_i-\vR_j )\cdot \va \, (\vR_i-\vR_j) \cdot \vb \Bigr\}^2.
\end{equation}
Since the Tb$^{3+}$ orbitals are aligned along $\va $, the exchange couplings both
parallel and perpendicular to the axis of the $f$-orbitals are negligible.
An exception to this conjecture is the exchange interaction
$J_{l4}^{xy}$ ($R=7.53$\,\AA).  Because this interaction span larger distances
than $J_1$ and $J_{l1}$, it may involve more complex exchange
pathways meditated by Ga ions, as discussed in Section V.  Although the $xy$ exchange
interactions can be FM ($J_3^{xy}$, $J_4^{xy}$, and $J_{l3}^{xy} >0$) or
AF ($J_2^{xy}$, $J_{l2}^{xy}$, and $J_{l4}^{xy} < 0$),
the largest intralayer and interlayer $xy$ exchange couplings
$J_2^{xy}$ and $J_{l2}^{xy}$ are both AF,
in agreement with the first-principles calculations discussed in the previous section.

Both the inelastic spectra and first-principles calculations indicate the the
exchange interactions in TIG are long-ranged.  Similar long-ranged interactions extending
over many Tb$^{3+}$ layers were found in the compounds TbNi$_2$Ge$_2$ and TbNi$_2$Si$_2$,
which display several magnetization steps and are possible examples of
``devil's staircases" \cite{blanco91, shigeoka92, budko99}.

Within the Kitaev model \cite{kitaev06} on a honeycomb lattice, strong SOC produces
a $J_{{\rm eff}}=1/2$ state and the exchange couplings on the three bonds of the non-distorted
honeycomb lattice are different. For TIG, the exchange couplings between Tb$^{3+}$ ions
in the distorted honeycomb lattice depend on the orientations of the coupled Tb 4$f$
orbitals.  This bond-dependent exchange is required to understand both the static {\it and} dynamic
properties of TIG.

As in other materials \cite{budko99, wiener00} containing Tb$^{3+}$ ions, the low-lying
crystal-field doublet in TIG affects the $\lambda $ anomaly of the
specific heat \cite{khan19b}, which exhibits an $R\,{\rm ln} 2$ entropy
characteristic of $J_{{\rm eff}}=1/2$ moments.  So there is no doubt that the
strong CF potential in TIG splits the $2J+1=13$ levels of Tb$^{3+}$ into a
low-lying doublet and 11 higher levels.

Of course, single-ion anisotropy (SIA) should vanish within the low-lying doublet
$\vert \Phi_{\pm }\rangle $ because $\langle \Phi_{\pm }\vert J_{\alpha }^2 \vert \Phi_{\pm }\rangle $
is the same for each state.  In the absence of easy-axis and easy-plane anisotropies,
a rigorous description of TIG must include 7 interaction terms per bond \cite{Li15, Maksimov19, Matsumoto20}:
isotropic Heisenberg exchange $J$, exchanges $J_x$ and $J_z$ coupling only the $x$
or $z$ spin components, symmetric exchange $J_{xy}$ and antisymmetric (DM) exchange $D$
coupling the $x$ and $y$ spin components, and finally, exchanges $J_{zx}$ and $J_{zy}$
coupling the $z$ and $x$ or $y$ spin components.
A complete model of TIG should contain at least five bonds: three bonds to produce
the two jumps in the magnetization with field along $\va $ and at least two additional
bonds between layers.  Adding three $\underline{g}$-tensor components but constraining the
antisymmetric $D$ exchange interactions using the observed canted moment,
a rigorous description of TIG then requires {\it at least} 37 parameters.
Needless to say, fitting 37 parameters is nearly impossible and defeats the
whole purpose of a model Hamiltonian. Therefore, we have used a phenomenological model containing SIA for
general spin $S$ with ``only" 13 terms to describe this system.
Aside from practicality, another advantage of this model is that the
exchange, anisotropy, and $g_{\alpha \alpha }$ components have direct physical interpretations.

The $\rm R_2T_3X_9$ family exhibits a  variety of ground states
that depend on the competition between long-range magnetic
interactions and magneto-crystalline anisotropy
arising from the interplay between
crystalline electric-field and Kondo effects.  Due to the large coordination
number ({\it i.e.}, the rare-earth R has 11 nearest-neighbor X-ligand atoms and
6 next-nearest-neighbor T-ligand atoms), a slight change in the local
environment surrounding the R atom (average bond-distance) can lead to
drastically different ground states ranging from a mixed-valent to a
Kondo-lattice system  \cite{dhar99,trovarelli99,okane02,niermann04}.
For example, $\rm Dy_2Co_3Al_9$ undergoes transitions into
two incommensurate states before locking into a low-temperature commensurate state \cite{gorbunov18}.
This complex phase diagram indicates significant magnetic frustration
due to the long-range exchange couplings which also appear in TIG.
However, the prevailing easy-plane anisotropy of TIG
drives the system into a commensurate spin state,
albeit one with many competing states of slightly higher energy.

To summarize, neutron diffraction and INS measurements were used to
investigate the  static and dynamical properties of the honeycomb-lattice
TIG.  Neutron diffraction measurements on a single crystal reveal a canted AFM
spin configuration with a moment of about 1.22$\mb $/f.u.~along $\vb$.
Fits to the inelastic spectrum indicate bond-dependent exchange interactions while
Monte-Carlo simulations and first-principles calculations suggest
competing ground states.  Consequently, TIG has a great
deal in common with other $J_{{\rm eff}}=1/2$ materials on a honeycomb lattice.

Research at ORNL's HFIR and SNS was sponsored by the Scientific User Facilities Division,
Office of Basic Energy Sciences, U.S.~Department of  Energy (DOE).
R.S.F., M.E.M., and D.P. acknowledge  support by the U.S. Department of Energy,
Office of Basic Energy Sciences, Materials Sciences and Engineering Division.
Work in the Materials Science Division at Argonne National Laboratory
(crystal growth and magnetic characterization) was supported by the U.S.
Department of Energy, Office of Science, Basic Energy Sciences,
Materials Science and Engineering Division.

\vfill
\bibliographystyle{h-physrev}

\end{document}